# Spatially-Resolved Characterization of Oil-in-Water Emulsion Sprays

Cheng Li,[1] Ruichen He,[1] Zilong He,[1] S. Santosh Kumar,[1] Steven A. Fredericks,[2] Christopher J. Hogan Jr.,[1] & Jiarong Hong[1,*]

[1]Department of Mechanical Engineering, University of Minnesota
[2]WinField® United
[*]To whom correspondence should be addressed: jhong@umn.edu

**Abstract:**

When an oil-in-water emulsion is utilized for application in a flat-fan spray, micrometer-sized oil droplets are thought to facilitate hole formation on the spray lamella, leading to an earlier breakup of the spray sheet and an increase in resulting droplet sizes. However, prior work has largely focused on changes to the mean or median droplet size, and has not examined the influence of oil-in-water emulsions on the complete droplet size distribution, as well as other droplet properties such as droplet shape and droplet initial velocity distributions in flat fan sprays. This study concerns the effects of pressure, spatial location, and application of oil emulsions on the resulting droplet size, eccentricity, as well as velocity distributions, all of which are crucial information in determining the dispersion dynamics of the droplets during the spray applications. Experiments were conducted with a TP6515 nozzle, with the abovementioned droplets information measured using digital inline holography (DIH). Results show that the DIH-inferred volumetric droplet size distributions (VDSD) span widely from sub-200 µm to over 2 mm in size. The application of an oil-in-water emulsion results largely in suppression of smaller droplets, while the VDSD is relatively insensitive to increasing the oil volume fraction beyond a critical level. DIH results are consistent with the observation that smaller ligaments are observed only in the breakup regions for the single phase water sprays, which breakup more violently and earlier than larger ligaments. It is these smaller ligaments that are suppressed in the oil-in-water emulsion, and which yield sub-millimeter droplets, some with sizes approaching sub-200 µm dimensions, corresponding to driftable sizes in agricultural settings. DIH additionally enables determination of size dependent droplet eccentricity and velocity measurements. Interestingly, the application of oil-in-water emulsion generally decreases the eccentricity, more significantly at the center than at the edge of the spray fan. We attribute this decrease to the increase in lamella sheet thickness and thus decrease in characteristic shrinkage rates, consistent with the observation using high speed shadowgraphs. In all instances, oil-in-water emulsion droplets have higher velocities than equivalent sized water droplets. We attribute this to the earlier action of the spray breakup process in the oil-in-water emulsion, reduced surface energy generation during breakup (larger droplets), and reduce energy dissipation during breakup with oil-in-water emulsions, leading to increased translational energy after the breakup process. Therefore, it appears that oil-in-water emulsion application simultaneously suppresses small droplet formation and increases droplet velocity, and hence spray penetration in agricultural application.

# 1. Introduction

Modulation of droplet size distributions in sprays is of crucial importance for a broad range of industries (Mugele and Evans, 1951; Lefebvre and McDonell, 2017). In particular, agrochemical application is frequently carried out via sprays, and the droplet size distribution of such sprays needs to be controlled carefully to avoid droplet entrainment in wind resulting in spray drift (Bilanin et al., 1989; Holterman et al., 1997; Reichenberger et al., 2007; Teske et al., 2009), while ensuring appropriate penetration of the spray into the crop canopy (Zhu et al., 2002; Sijs and Bonn, 2020) and plant wetting (Lake, 1977). One widely applied spray drift mitigation strategy involves utilizing a two-phase, oil-in-water emulsion as an adjuvant towards producing larger droplets or mitigating fine droplet formation in sprays (FOCUS, 2007; Hoffmann et al., 2008; Qin et al., 2010). A detailed understanding of dilute oil-in-water emulsions spray breakup and the resultant droplet size distribution is therefore critical in improving agrochemical spray efficacy and towards spray drift reduction.

Droplet formation in single-phase water sprays is understood to be dominated by two processes, namely, ligament formation brought about by Squire wave development (Squire, 1953; Dombrowski et al., 1954; Fraser et al., 1962; Dombrowski and Johns, 1963; Villermaux and Clanet, 2002; Altieri et al., 2014; Kooij et al., 2018) and subsequent ligament breakup due to capillary instabilities (Rayleigh, 1878). Conversely, the inhomogeneity in oil-in-water emulsion sprays leads to perforations (hole formation) on the liquid sheet prior to ligament formation (Dombrowski et al., 1954), which increase the characteristic thickness of the liquid sheet upon breakup into ligaments, presumably generating coarser spray droplets (Ellis et al., 1997; Dexter, 2001; Qin et al., 2010; Hilz et al., 2012; Altieri et al., 2014; Cryer and Altieri, 2017). A substantial body of work has been devoted to understanding the effects of oil-in-water emulsions on hole formation in a spray sheet, as well as to quantifying resultant integral metrics of droplet size, e.g. volumetric median diameter (VMD). For example, Dexter (2001) and Qin et al. (2010), used high speed imaging and laser diffraction spray analyzer to investigate the effects of oil-in-water emulsion on the spray characteristics by flat fan nozzles. They both suggested that the stretching of the emulsion droplets inside the lamella presents localized perturbations for holes to form. Based on laser diffraction size results for various oil-in-water emulsion sprays, finding that oil with a high positive spreading coefficient results in coarser sprays, Hilz et al. (2012) later proposed that oil droplets migrate to and finally spread on the air/water interface, generating a secondary flow, which acts as a perturbation and causes hole formation. Vernay et al. (2015) measured the thickness and expansion rate of holes in a model experiment involving liquid sheets with oil droplets of volume median diameter of ~ 20 μm. In line with the propose mechanism from Hilz et al. (2012), they identified a pre-hole process whose growth dynamics matches that of liquid spreading on another liquid with higher surface tension.

Despite such progress in describing oil droplet spreading, hole formation, hole growth, and ligament breakup mechanistically, the aforementioned studies do not fully quantify the effect an oil-in-water emulsion has on droplet size distributions, including not only the mean size, but also the shape of the distribution and its polydispersity. Further research is thus needed to describe how full droplet size distributions are influenced by spray operating conditions and spray composition for a wide variety of spray and nebulizer systems. One of the major reasons improved droplet size distribution characterization remains necessary is that there are challenges in complete characterization of spray droplet size distributions via conventional in-situ droplet size distribution diagnostic techniques, i.e. laser diffraction and phase doppler analysis, which are typically calibrated with monodisperse standards, and not tested in their ability to infer distribution shapes.

For this reason, the present study applies Digital Inline Holography (DIH) to measure droplet size distributions, joint size-eccentricity and size-velocity probability density functions (pdfs), and size-dependent droplet flux distributions resulting from flat fan water and oil-in-water emulsion sprays. Holography is a three-dimensional, direct imaging technique, which consists of recording the interference patterns between a reference beam and light scattered by individual objects in the volume of interest, with orders of magnitude better depth of field than conventional photography (Vikram, 1992; Hariharan and Hariharan, 1996). DIH is the simplest form of holography and has been widely used to measure the 3D location, motion, and size of particles, droplets, bubbles, and microorganisms in numerous applications (Katz and Sheng, 2010). In particular, DIH has been applied to measure size distributions of bubbles (Katz, 1984; Shao et al., 2019), oceanic sediment (Graham and Nimmo Smith, 2010), cloud droplets (Fugal and Shaw, 2009; Beals et al., 2015), oil droplets in emulsions (Tian and Barbastathis, 2010; Li et al., 2017), and droplet fragments from breakup of larger drops (Guildenbecher et al., 2017). Compared to laser diffraction, phase doppler analysis, and traditional imaging, a greater amount of information is captured by DIH (Kumar et al., 2019; Kumar et al., 2020). DIH can provide three-dimensional data for the location, size, and velocity of individual droplets, simultaneously, and can be used to determine size distributions and pdfs without assumptions on particle shape (Guildenbecher et al., 2017). Using DIH, our goal in the current study is to precisely determine size distributions of droplets produced in flat fan sprays single-phase water and two-phase oil-in-water emulsions, in an effort to better quantify how changes in the microscopic spray breakup process ultimately influence droplet size distributions, at different locations in the spray. The facility, oil properties, and measurement techniques are described in section 2, followed by the results in section 3 and conclusion in section 4.

## 2. Experimental Methods

### 2.1. Spray and Wind Tunnel Operation

Measurements were carried out using a spray boom-equipped recirculating wind tunnel facility at the University of Minnesota. A schematic diagram of the wind tunnel is provided in Fig. 1a. The length of the tunnel (spanning from left to right in the diagram) is 6.8 m, the width (bottom to top in the diagram) is 3.8 m, and the height is 2.7 m. The test section of the tunnel, shown at the bottom of the schematic diagram, is a glass wall region which is 0.91 m in width, 3.20 m in length, and 1.83 m high. The test section is equipped with a single-nozzle spray boom which can be automatically adjusted in vertical position, with automated control of the supply pressure ($p$) and measurement of the mass flow rate ($q$) via a Badger Coriolis mass flow meter (RCT 1000). We examined sprays produced by a TP6515 (Teejet Technologies) flat fan nozzle, operated at liquid upstream pressures of 0.69 bar and 2.76 bar with the laboratory temperature near 20º C. The flow rate is tested to be dependent on the pressure ($q \sim p^{1/2}$). Both tap water and an oil-in-water emulsion of cod liver oil (fish oil, Twinlab) at oil volume percentages of 0.1%, 0.2%, and 1% were used in spray generation. The water and emulsion properties (density, kinematic viscosity, interfacial tension with water, and interfacial tension with air) are shown in Table 1. Values are taken from prior studies wherein viscosities were measured using Cannon-Fenske opaque viscometers while the surface tensions and interfacial tensions were determined using the pendant drop method (Song and Springer, 1996; Murphy et al., 2015). Because fish oil has a relatively low interfacial tension with water and low surface tension with air relative to water, its spreading coefficient (the surface tension of water minus the fish oil surface tension and interfacial tension) is relatively high. Oil-in-water emulsions were prepared using Oster 700 W Blender operated at the highest power setting for 1 minute, followed by immediate spray application.

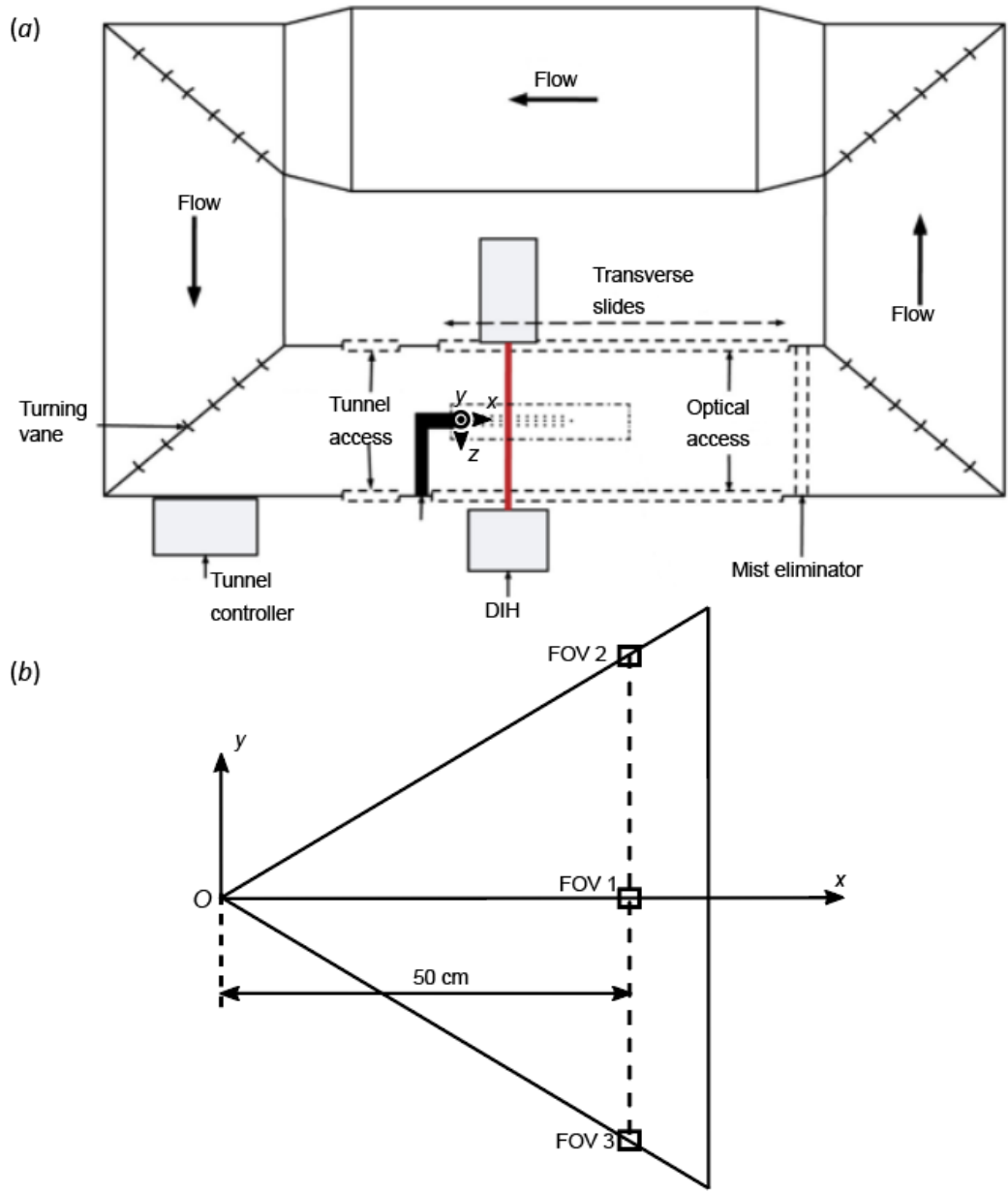

**Fig. 1. (a)** A schematic diagram of the recirculating wind tunnel used for DIH measurements of flat fan sprays. Air flow was not utilized in the current study, but the complete position of the spray boom and digital inline holography (DIH) system are shown to make clear how measurements were performed. **(b)** A side view diagram of the DIH measurement locations in the flat fan spray.

**Table 1.** Properties of water and fish oil at 20° C. $\rho$: density. $v$: kinematic viscosity. $\sigma_w$: interfacial tension with water. $\sigma_a$: interfacial tension with air.

| Fluid | $\rho$ (g/m³) | $v$ (cSt) | $\sigma_w$ (mN/m) | $\sigma_a$(mN/m) |
|---|---|---|---|---|
| Water | 1.00 | 1.0 | NA | 73 |
| Cod liver Oil | 0.92 | 63.1 | 14.9 | 22.5 |

*2.2. Digital Inline Holography & High-speed Shadowgraph Imaging*

During spray operation, DIH measurements were made at the center (FOV 1) and two periphery (FOV 2 and 3) locations within the spray fan, which is depicted in Fig. 1b. DIH was carried out using a high-speed camera (Phantom v710), and an imaging lens (Nikon 105 mm f/2.8) with laser illumination via a 12 mW helium-neon laser (REO Inc.). Laser intensity was controlled via a neutral density (ND) filter (Thorlab Inc.), and laser spatial coherence was improved via spatial filter (Newport Inc.). The laser, filters, and a collimation lens with 75 mm focal length (Thorlabs Inc.) were mounted on an optical breadboard and directed at the imaging lens via two turning mirrors, resulting in a 50 mm diameter collimated Gaussian laser beam. For lower pressure measurements, DIH image processing, to first identify droplets, then to correct for occluding droplets and determine droplet diameters (modeling droplets as ellipses and determining projected area equivalent diameters, as well as aspect ratios), was conducted using the methods described in Kumar et al (2019). At higher pressure, due to the higher droplet number density, the conventional segmentation method proves incapable of extracting droplet size and location. Therefore, a novel machine learning approach was utilized to process the DIH data at higher pressure (Shao et al., 2020). Specifically, by using a modified U-net architecture with three input channels including original holograms, holograms reconstructed at each longitudinal location, and minimum intensity projection along longitudinal direction, two output channels are predicted containing reconstructed objects and their centroid information at each longitudinal location separately. The model is pretrained using synthetic holograms that have similar characteristics as the real holograms (polydisperse near-spherical droplets). For both size and location output channels, the binary cross entropy loss function is implemented. During training, a set of real reconstructed holograms with the manual extracted ground truth is used as a validation to determine which trained model to use for the hologram processing task. DIH was calibrated following the method introduced in Kumar et al (2019) via examination of a precision microruler (Thorlabs) and measurement of monodisperse droplet generated by a vibrating orifice aerosol generator (Berglund and Liu, 1973). The DIH system utilized has a dynamic range from 21.3 μm (a single pixel) to 10.9 mm (512 pixels), covering the range of droplet sizes of interest.

In addition to droplet size measurement, DIH was utilized to determine droplet speeds in the streamwise (axial) direction at the measurement location. To do so, we utilized the high-speed camera's Memgate function with a frequency of 100 Hz and 4% duty cycle. Because the camera has an internal data acquisition rate of 25,000 frames/s, this led to 100 independent 10-frame samples every second, wherein different droplets were observed in each sample. Using these data, the velocimetry algorithm of Ouellette et al (2006) was used to extract Lagrangian tracks for observed droplets. The tracking algorithm was applied to droplet centroids, which were evaluated during droplet size and shape extraction over numerically reconstructed holograms, using Matlab. Each droplet trajectory consisted of a maximum of 10 frames, and droplet streamwise direction speeds were taken as the average speed of the centroid over the observed frames. All data processing was carried out at the Minnesota Supercomputing Institute, and for each measurement location, measurements led to mean droplet diameters, eccentricities (specifically the first eccentricity, spanning from 0 for a circle to 1 for a straight line), and speeds for each observed droplet. Data were then binned to generate one-dimensional size distribution functions reported on a volumetric basis, size dependent eccentricity distributions, and joint size-velocity pdfs. For all sprays, size distributions are reported at $x = 50$ cm ($x/D_N = 122$) downstream of the nozzle outlet, where $D_N$ corresponds to the width of the nozzle orifice. All measurements were centered about a hologram focal plane at $z = 0$ (the center of the spray in the direction of the laser beam) and were digitally reconstructed at multiple planes separated by 0.5 mm across a depth of 20 cm, to identify droplet focal planes along the spray depth. Since the flat spray also expands significantly over the depth direction (Dexter (2001) as droplet/ligaments ejects at angles during wave growth and sheet breakup, identification of droplets at extended depth is critical in correctly measuring the droplet size distributions. All detected droplets are finally presented *x-y* plane, regardless of depth values (*z*) forming depth-compressed results. The depth-compressed sample trajectories are provided in Fig. S1 of the supporting information and a depth compressed sample video is also provided as supporting files. We specifically probe the influence of fish oil volume fraction on the size distribution at the *y* = 0 location. Subsequently, for water and the 0.1% oil-in-water emulsion spray, we report size distributions and joint pdfs at variable *y* locations, spanning to *y* = ± 30 cm from the spray centerline. Table S1 of the supporting information summarizes the complete number of samples collected and the number of holograms used for each sample. For joint pdfs and the resulting flux distributions, we utilized two 30-s collection periods, leading to 6,000 independent hologram measurements.

In addition to DIH measurements, to qualitatively examine spray breakup, we conducted shadowgraph imaging using the high-speed camera and imaging lens to collect time- resolved shadowgrams at the exit of the spray nozzle. Furthermore, the identified starting point of lamella breakup both at the center and edge are scrutinized by time resolved cinematic DIH. We report on shadowgram and cinematic DIH results first prior to DIH measurements on droplet size, shape, and velocity to demonstrate the spray breakup into droplet occurred upstream of the measurement location in all conditions.

## 3. Results

### 3.1. Qualitative Examination of Spray Breakup Process

Sample shadowgrams at both examined liquid supply pressures and for water and oil-in-water emulsions are shown in Fig. 2. Focusing first on the lower liquid supply pressure series of images (a,c,e), shadowgrams confirm the formation of holes in the oil-emulsion spray fan which do not appear in the single phase water spray fan. This is consistent with prior studies of spray fan breakup in oil-in-water emulsions (Qin et al., 2010; Hilz et al., 2012; Altieri et al., 2014; Vernay et al., 2015; Vernay et al., 2016; Vernay et al., 2017). Increasing the oil volumetric concentration in the emulsion visibly leads to initial hole formation closer to the spray nozzle and an apparent increase in the hole formation rate (number of holes per image). Hole formation is also evident in shadowgrams collected for the higher-pressure case. However, as higher backing pressure leads to smaller wavelength oscillations in the sheet, holes are more difficult to discern in high pressure spray shadowgrams. Consistent with the depiction provided by Vernay et al (Vernay et al., 2015; Vernay et al., 2017), holes visibly expand as they migrate downstream in the spray fan, and upon merging, they lead to ligament formation in the fan. Ligaments, in turn, break-up into droplets. From visual observation and in-line with prior work, it appears that hole formation leads to a distinct change in droplet formation in two-phase fluid sprays.

The breakup region, namely the region where lamellae break and ligament formation and breakup occur, is visually identified using shadowgraph imaging results. Accordingly, the 0.1% oil-in-water emulsion spray breakup region is ~ 1.7 cm upstream at the centerline and ~ 3.1 cm upstream at the edge of single-phase water spray breakup region, at 0.69 bar, with similar upstream breakup regions observed for the other examined operating conditions. Sample time-resolved microscopic views of the breakup region from holographic images (marked as the red squares in Fig. 2) are provided for both water and 0.1% oil-in-water emulsion cases in Fig. 3 and 4 respectively, with each showing both the breakup processes at the spray center and edge. For the single-phase water spray fans (Fig. 3), over the course of image collection, the lamellae within the spray fan are observed to form complicated ligaments networks of variable sizes. The smaller ligaments break up earlier, presumably due to increased susceptibility to perturbations. Breakup forms a large number of sub-millimeter droplets, some with sizes approaching sub-200 μm dimensions, corresponding to driftable sizes in agricultural settings (Zhu et al., 1994; Guler et al., 2007)). Meanwhile, the larger ligaments continue to evolve, with their diameters modulated by the surrounding flow and neighboring ligaments. Larger ligament breakup can be partially observed later in the time series (Fig. 3a at 0.68 ms and 3b at 0.72 ms). Comparatively, the droplets released from the larger ligaments appear to be larger, with fewer falling into the sub-200 μm size range.

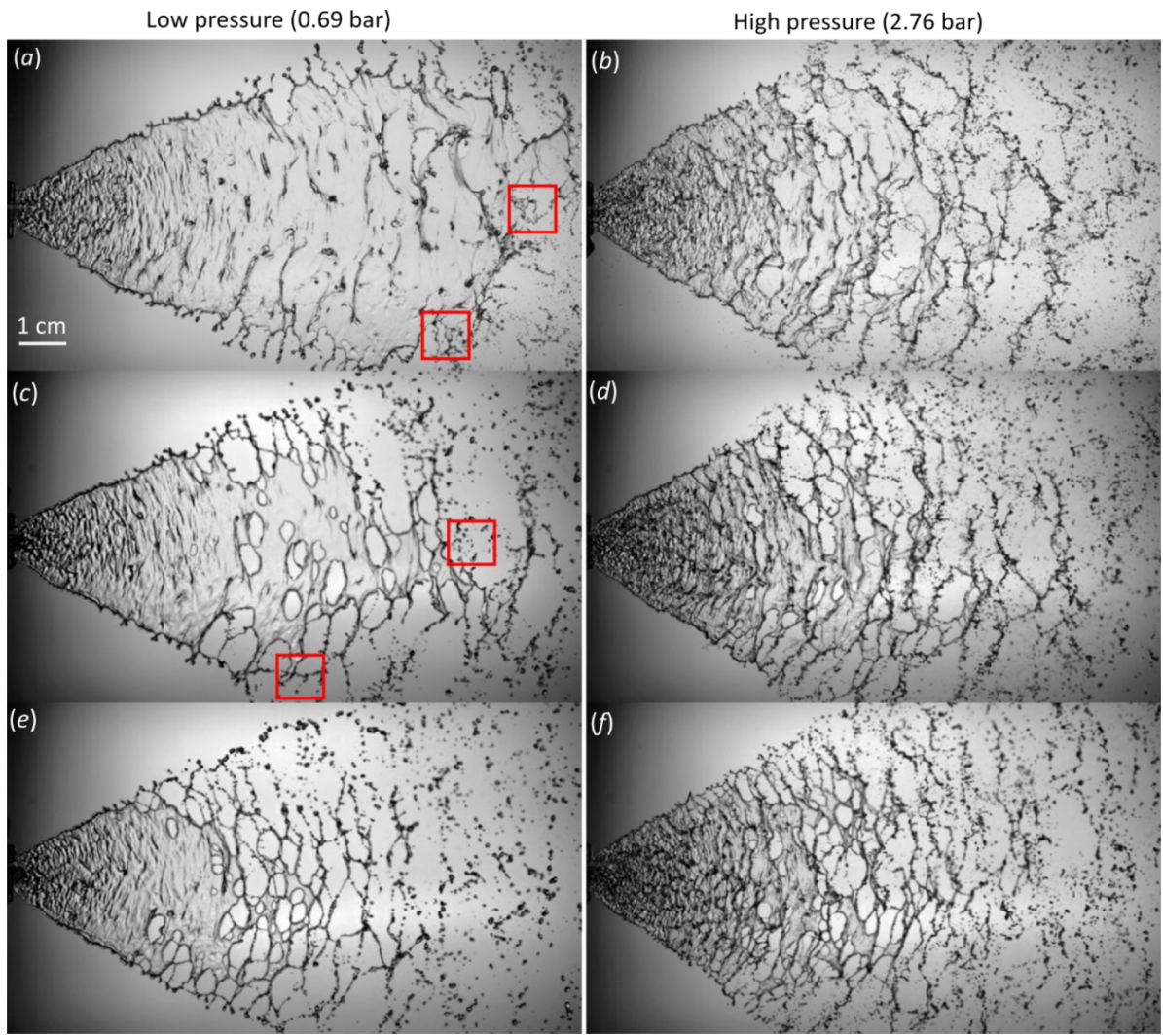

**Fig. 2.** Sample high speed shadowgraphs of water and oil-in-water emulsion sprays from a TP61515 flat fan nozzle with liquid supply pressures of at 0.69 bar (a,c,e) and 2.76 bar (b,d,f). (a,b) water; (c,d) 0.1% oil-in-water emulsion; (e,f) 1% oil-in-water emulsion. The red squares correspond to high speed DIH field of views for results reported in Fig. 3.

Visualization of the oil-in-water emulsion spray breakup region (Fig. 4 ) suggests that hole formation leads to formation of thicker ligaments than the ligaments in single-phase spray fans, as there are fewer small droplets observed, they are less likely to disintegrate into smaller droplets. To be consistent in comparison, we do not determine droplet size distributions in the abovementioned regions since the initial breakup processes are still in progress. Instead, as shown in the next section, droplet size distributions are presented at $x$ = 50 cm downstream, where all breakup events have ceased for all tested cases.

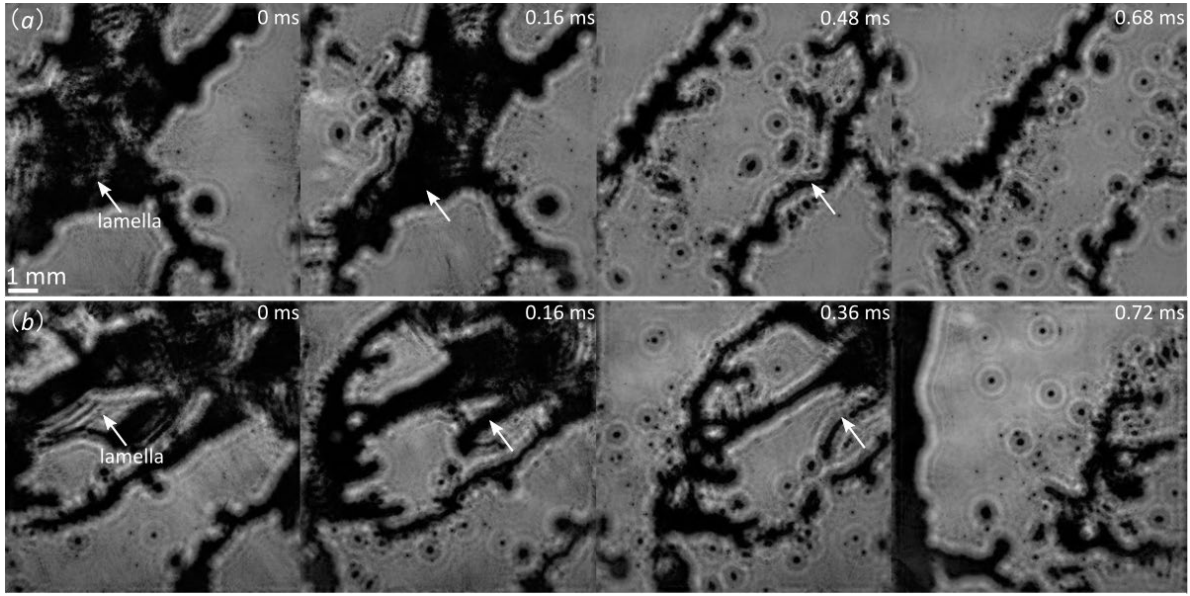

**Fig. 3.** Time series of depth-compressed holographic images of water spray at 0.69 bar liquid supply pressure revealing ligament and lamella breakup. (a) at the spray center: $x$ = 11.3 cm, $y$ = 0; (b) at the spray edge: $x$ = 9.1 cm, $y$ = 2.5 cm. White arrows indicate lamellae and ligaments observed in both single-phase and two-phase sprays.

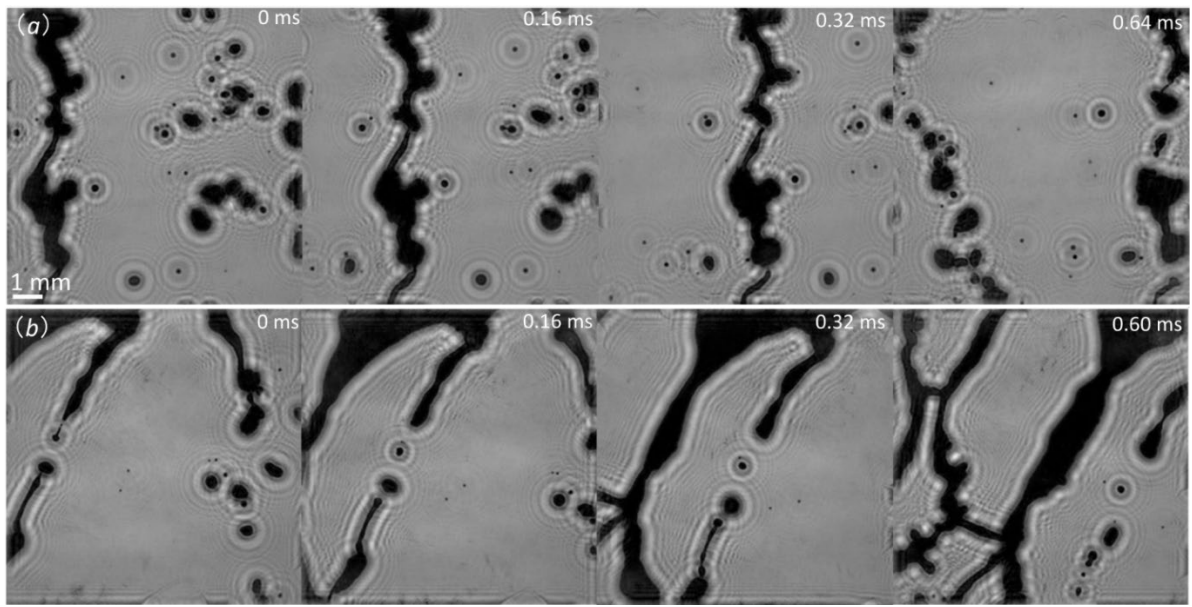

**Fig. 4.** Time series of depth-compressed holographic images of 0.1% oil-in-water emulsion spray at 0.69 bar liquid supply pressure. (a) at the spray center: $x$ = 9.6 cm, $y$ = 0; and (b) at the spray edge: $x$ = 6.0 cm, $y$ = 2.8 cm.

## 3.2 Droplet Size Distribution Functions

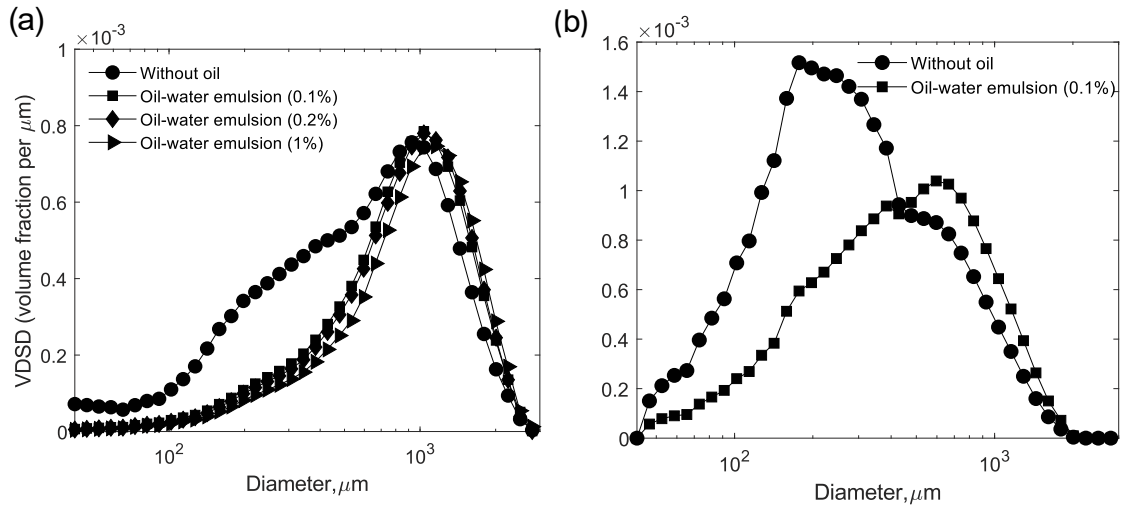

**Fig. 5.** Normalized volumetric droplet size distributions at $x = 50$ cm, $y = 0$ cm, for single phase water sprays and variable concentration two-phase oil-in-water emulsions. (a) 0.69 bar liquid supply pressure. (b) 2.76 bar liquid supply pressure.

While imaging techniques allow for a qualitative description of hole formation, and sheet and ligament breakup, of greater interest is how the volumetric droplet size distribution (VDSD) resulting from the spray is affected by the two-phase emulsion, as spray performance is largely governed by the VDSD. To first examine differences in single-phase and two-phase emulsion spray-generated droplets, we plot VDSDs derived from DIH measurements at the $x = 50$ cm, $y = 0$ location for both liquid supply pressures in Fig. 5. In line with expectations from visual observation, small droplet formation appears to be depressed in two-phase emulsion sprays, when compared to single phase water sprays. At the same time, the VDSD is relatively insensitive to oil volume fraction, and there is little difference in VDSD brought about by increasing the volume fraction of oil beyond 0.1%. Such observation is consistent with previous study by Vernay *et al.*(2016) A summary of basic VDSD properties is provided in table 2. For the water sprays, with increasing the supply pressure, the integral droplet size generally follows the power law dependence reported by Kooij *et al.*(2018). Examination of the $d_{10}$, $d_{50}$, and $d_{90}$ results (defined in the table caption) quantitatively shows that the key difference between two-phase emulsion sprays and single-phase sprays is an elevated $d_{10}$ for two-phase emulsions, and correspondingly a decrease in the distribution span. Increases volumetric mean diameters are also evident in emulsion sprays. The net effect of the emulsion, in terms of suppressing small droplet formation and decreasing the span of the VDSD is reduced at the higher pressure. At 0.69 bar, the 0.1% oil emulsion spray yields a 20% span decrease and 69% mean diameter decrease. However, at 2.76 bar the emulsion yields only a 14% span decrease and 47% of volumetric mean diameter decrease. The decrease in effectiveness with increasing pressure in span suppression and integral droplet size increase is associated with the inherent earlier breakup at higher pressure, even for water spray. With increasing pressure, for oil-in-water emulsion sprays, the oil droplet has less time to induce perforations compared to lower pressure case. Thus, fewer perforation events occur before the breakup. This argument is consistent with the high-speed visualization in Fig. 2. Moreover, the centerline DIH results are in qualitative agreement with the centerline results of Hilz *et al.* (2012) for an XR11003 nozzle at 3 bar liquid supply pressure, inferred from laser diffraction measurements, though Hilz et al. (2012) only report VMDs (medians) and cumulative fractions of distributions below 100 μm.

**Table 2.** A summary of centerline volumetric droplet size distribution properties, as determined via digital inline holography (DIH). $d_{10}$: droplet diameter corresponding to 10% of the cumulative distribution. $d_{50}$: droplet diameter corresponding to 50% of the cumulative distribution. $d_{90}$: droplet diameter corresponding to 90% of the cumulative distribution. Δ: distribution span, $(d_{90}-d_{10})/d_{50}$ (Lefebvre and McDonell, 2017). Mean: volumetric mean diameter.

| Pressure/ bar | $|y|$/cm | oil Vol% | $d_{10}$/μm | $d_{50}$/μm | $d_{90}$/μm | Δ | Mean/ μm |
|---|---|---|---|---|---|---|---|
| 0.69 | 0 | 0 | 371 | 994 | 1713 | 1.35 | 295 |
| 0.69 | 20 | 0 | 725 | 1305 | 1845 | 0.75 | 429 |
| 0.69 | 0 | 0.1% | 575 | 1147 | 1817 | 1.08 | 499 |
| 0.69 | 20 | 0.1% | 744 | 1386 | 2077 | 0.93 | 681 |
| 0.69 | 0 | 0.2% | 594 | 1173 | 1815 | 1.04 | 518 |
| 0.69 | 0 | 1% | 645 | 1237 | 1902 | 1.02 | 547 |
| 2.76 | 0 | 0 | 168 | 497 | 970 | 1.62 | 201 |
| 2.76 | 30 | 0 | 225 | 533 | 874 | 1.21 | 243 |
| 2.76 | 0 | 0.1% | 255 | 680 | 1072 | 1.20 | 272 |
| 2.76 | 30 | 0.1% | 250 | 618 | 960 | 1.15 | 267 |

More detailed quantification of the difference between two-phase emulsion sprays and single-phase water sprays is possible with DIH by examining VDSDs at variable vertical positions. As an example, Figure 6 focuses on location effects on VDSDs at the spray center and edges, for water and 0.1% oil-in-water emulsion spray with varying pressure. As expected, all VDSDs are broadly distributed from below 100 μm to above 2 mm, with each size distribution affected by pressures, location, and spray liquid properties.

The size distributions at the spray edge are distinct from corresponding distributions at the spray center, particularly at the lower test pressure. The volume fraction of droplets less than 100 μm in diameter at the spray edge is smaller than that at the spray center, regardless of the operating pressure and emulsion concentration. Although at 0.69 bar (Fig. 6a), size distributions are shifted to larger sizes at the spray edge for both water and the oil-in-water emulsion, shifts are less pronounced at 2.76 bar (Fig. 6b). Such observation can be further evident from the characteristic droplet sizes displayed in Table 2. For example, at lower pressure, the volumetric mean diameter increases 45% and 36% from center to edge, for water and 0.1% oil-in-water emulsion sprays, respectively. The increase in volumetric mean diameter is reduced or even reversed at the higher pressure, with 21% increase for water spray and a 2% decrease for 0.1% oil-in-water emulsion spray. The increase of characteristic droplet size at the edge of the spray has been observed in previous studies (Hilz et al., 2012; Vernay et al., 2016), and is supported by prior observations. Specifically, it is believed to be associated with a thicker rim developed at the periphery of the spray due to surface tension. This rim is clearly seen in Fig. 2 in all examined instances. However, it is less prominent at the higher pressure due to increased spray inertia compared to surface tension forces. The estimated rim thicknesses are ~0.8 mm and ~ 0.4 mm near the nozzle, at lower and higher pressure, respectively. The finding that the addition of oil diminishes spatial effects is attributable to the formation of rims within the spray interior, i.e. hole formation leads to thicker interior rims, as observed in Fig. 2c-f. Measured by Vernay *et al*. (2015), a hole's rim size can be over two times the size of the mean sheet thickness. Consequently, for oil-in-water emulsion sprays, subsequent breakup in the center tends to generate similar droplet size distribution to the edge.

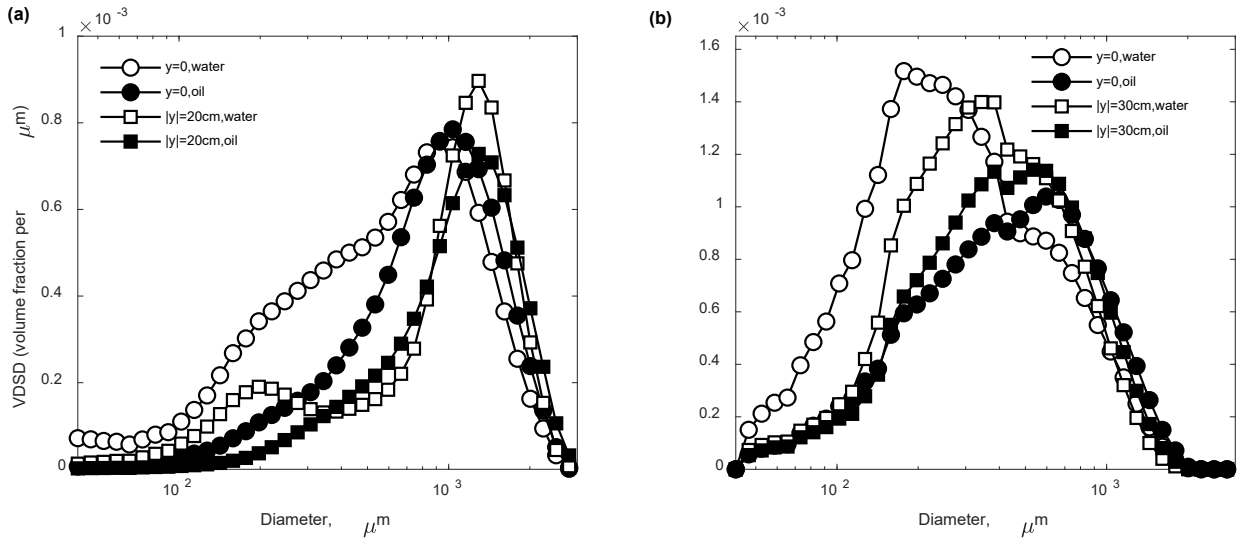

**Fig. 6.** Normalized volumetric droplet size distributions measured at both spray center and edge, at $x = 50$ cm. (a) water and 0.1% oil in water spray at 0.69 bar; (b) water and 0.1% oil-in-water emulsion spray at 2.76 bar.

*3.3 Eccentricity and velocity of droplets*

Beyond detection and characterization of droplet size, our initial reports (Kumar et al., 2019; Kumar et al., 2020) on DIH analysis of flat fan sprays demonstrate that DIH can also quantify shape of spray droplets in terms of the eccentricity of the droplets. The eccentricity is measured by fitting an ellipse that best fits the droplet image, the eccentricity, $e = (1-b^2/a^2)^{1/2}$, where $a$, and $b$ correspond to the long and short axis of the fitted ellipse. We present mean centerline and edge eccentricities ($<e>$) in Fig.s 7a &b respectively using the same logarithmic bin sizes as the VDSDs presented, but only size bins with at least $10^5$ particles to ensure statistical significance for the resulting mean eccentricity values. Additionally, as the measurement of small droplets eccentricity is limited with current measurements and analysis algorithms (Kumar et al., 2019), only eccentricity values for droplets larger than 300 μm is reported. In both the low and high pressure sprays, and for water and oil-in-water sprays, there is a positive correlation between diameter and eccentricity, with eccentricity exceeding 0.5 for a large fraction of the droplets larger than 1 mm. Meanwhile the operating pressures, locations in the spray, and use of an oil-in-water emulsion all modulate the mean eccentricity profile. First, as expected, the increase of pressure also increases the mean eccentricity of droplets for all droplet sizes of both water and 0.1% oil-in-water emulsion sprays at both the edge and the center of the spray. Second, the use of an oil-in-water emulsion generally decreases the eccentricity, more significantly at the center than at the edge. To clarify this finding, Fig. 8 compares the sheet breakup time series of water and oil-in-water emulsion sprays and quantifies the characteristic liquid sheet shrinkage rates ($u_s$) until breakup. For the latter, we use the measured sheet width ($w$) and time change between images, defining the sheet shrinkage rate $u_s = \Delta w/\Delta t$. The first two panels of Fig. 8a and 8b show different stages of the sheet shrinkage with the sheet width measured, while the last panel depicts an instance when two rims hit one another. The red arrow indicates the location where the sheet width is measured in previous two panels. For the water case, no clear ligaments are observed. In fact, multiple small fragments of ligaments and droplets are observed, indicating a large shrinkage rate. For the oil-in-water case, however, a single coherent ligament is formed. It is later elongated and breaks up into droplets due to capillary instability. The resulting shrinkage rates calculated are provided in Fig. 8c. As expected, the sheet shrinkage rate increases over time, due to the acceleration driven by surface tension for both cases. During the final stage of ligament formation, the characteristic sheet shrinkage rate is roughly one order of magnitude different between water and oil-in-water sprays. The large difference in sheet shrinkage rates may be associated with the reduced breakup peripheral length scale at the initial stage of the liquid sheet shrinkage and thus less pronounced surface tension effects due to earlier lamella breakup with the addition of oil emulsion.

As a result, breakup in oil-in-water emulsions is less likely to lead to violent breakup (c.f. Fig. 3.) resulting in internal velocity differences within droplets. As it is such velocity differences which lead to droplet oscillation, rotation, and increased eccentricity, oil-in-water emulsion use results in smaller mean eccentricity for all corresponding size bins shown in Fig. 7. Finally, the mean droplet eccentricity is lower at the spray edge than that at the spray center. For example, the overall mean eccentricity for droplets with $d > 300$ μm, the mean eccentricity at the spray center is 22%, and 9% larger than that at the spray edge at 0.69 bar, for water and 0.1% oil-in-water emulsion sprays, respectively. The corresponding values changed to 8% and 16% respectively at 2.76 bar.

Our results suggest that flat fan spray-generated droplets produced in the examined size range are best described non-spherical "packets" of liquid which are oscillating and rotating (Kumar et al., 2020), as opposed to static spherical droplets. Such information is valuable in more accurately modeling the transport of spray droplets and spray drift. Conventional spray drift simulation studies assume spherical droplets in calculating drag force, settling velocities, and droplet response time (Miller and Hadfield, 1989; Holterman et al., 1997). However, non-spherical droplets can display more complicated motion than spherical droplets (Haider and Levenspiel, 1989; Tran-Cong et al., 2004; Bagheri and Bonadonna, 2016), which ultimately may need to be concerned in improving modeling of droplet transport in sprays.

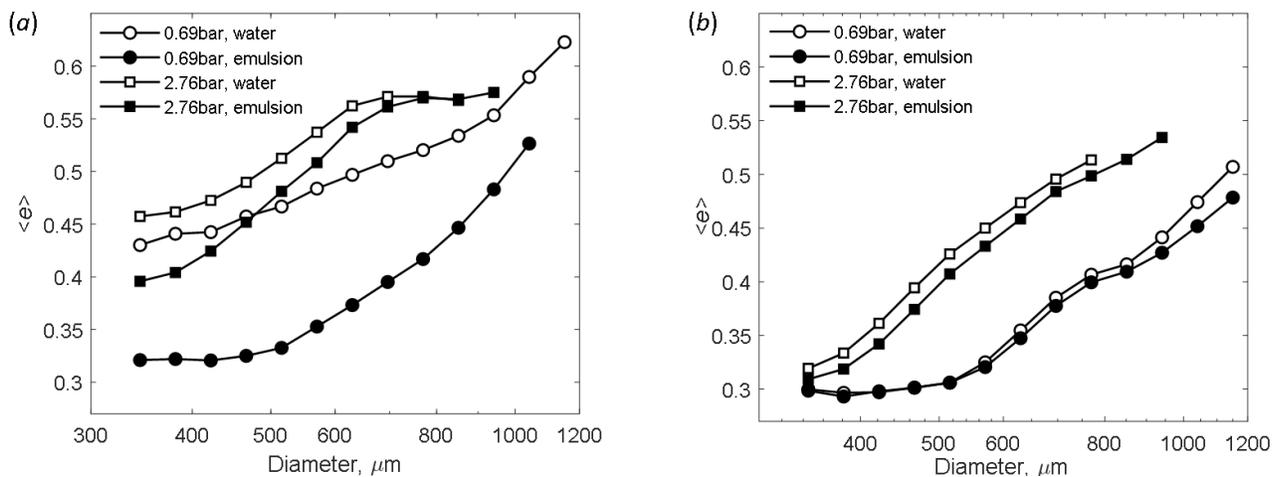

**Fig. 7.** Effects of pressure and oil-in-water emulsion on the mean droplet eccentricity at the spray center ($y = 0$) (a). and at the spray edge ($|y| = 20$ cm for 0.69 case and $|y| = 30$ cm for 2.76 bar case) (b). All emulsions are 0.1% oil by volume. All emulsions are 0.1% oil by volume.

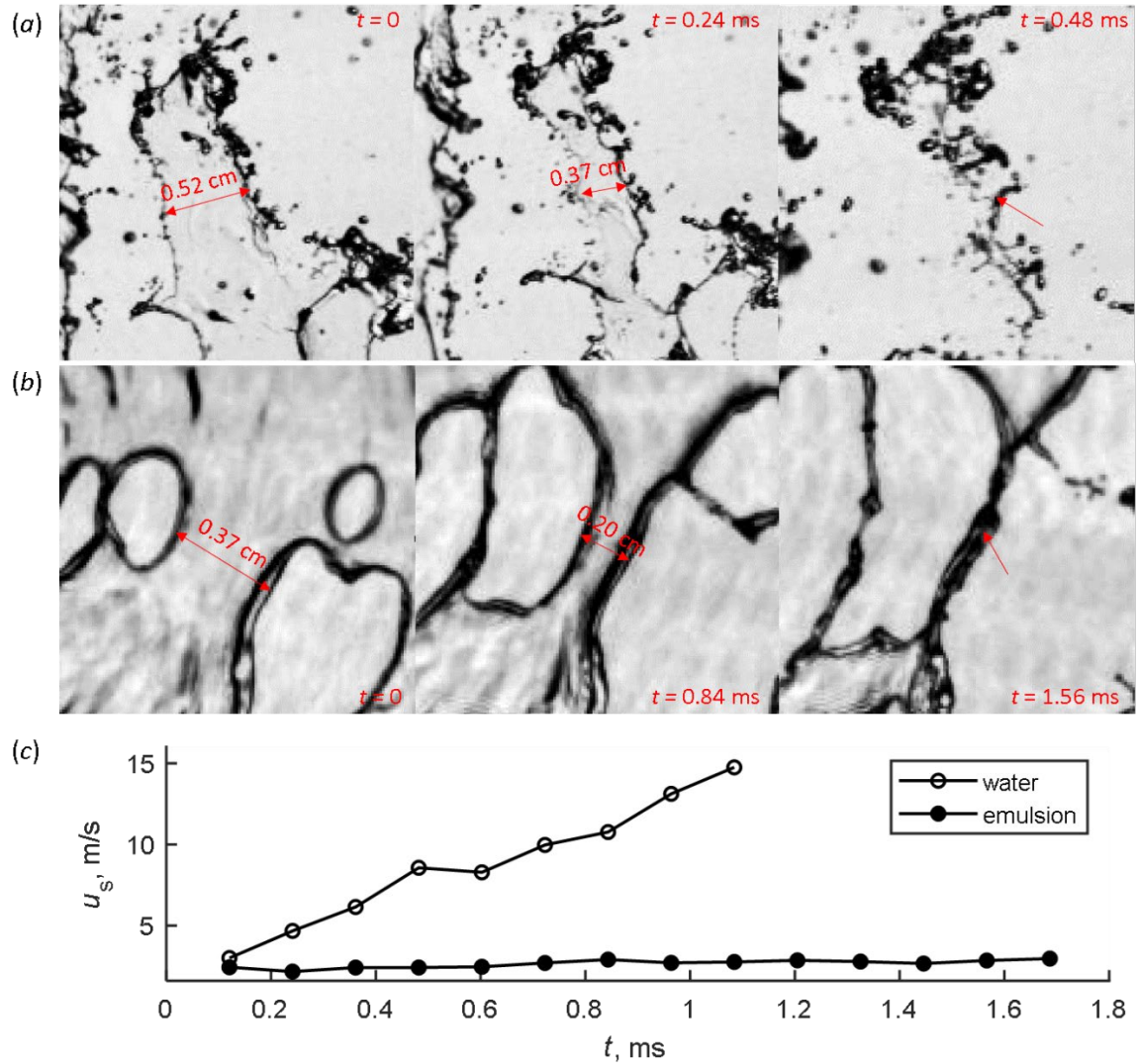

**Fig. 8.** Sample time series showing the spray sheet breakup at 0.69 bar liquid supply pressure for (a) water spray and (b) 0.1% oil-in-water emulsion spray. The breakup events are captured at $x = 11.74$ cm, $y = 0$, and $x = 5.61$ cm, $y = 0$ respectively. (c) Temporal evolution of sheet shrinkage rate.

We also conducted investigation on the effects of oil-in-water emulsion and pressure on spray droplet velocity. Fig. 9 shows a comparison of the mean droplet axial velocity ($<u>$, averaged over all droplets in each size bin) under the same conditions as the eccentricity results. For all examined instances, at small droplet size the diameter *vs* axial velocity curves display linear-log scaling, i.e., $<u>$ increases almost logarithmically with the increase of droplet size at small sizes, but then levels off to a region where $<u>$ varies little with increasing size. These curve shapes are qualitatively consistent with expectations from the standard drag curve for spheres, which leads more drag per unit mass on smaller drops. The drag coefficient of drops follows $C_D \sim 1/Re_D$ for small droplets of small $Re_D$, with $Re_D$ as the droplet Reynolds number following the standard definition, and $C_D = 2F_D/(\rho u^2 d^2)$ where $F_D$ is the drag force on the droplet, $\rho$ is air density, and $d$ the droplet diameter. $C_D$ gradually evolves to a constant for larger droplets with larger $Re_D$, and there is a transition regime in $C_D$ between these two limits. Examining the measurement result, for the 0.69 bar water case, ~6 m/s, 300 μm droplets and ~9 m/s 1000 μm droplets result in $Re_D$ of 118 and 593, respectively, where the drag force ($F_D$) falls into this transition regime. Consequently, the drag force varies as $F_D \sim d^x$ with $1<x<2$. The magnitude of the acceleration of a droplet is $a \sim F_D/d^3$ and thus $a \sim d^{x-3}$. Conversely, in the large diameter limit, $C_D$ is a constant, and $a \sim F_D/d^3 \sim d^{-1}$. The decrease of the magnitude of acceleration for small droplets ($\sim d^{x-3}$) is thus faster than the larger droplets ($\sim d^{-1}$).

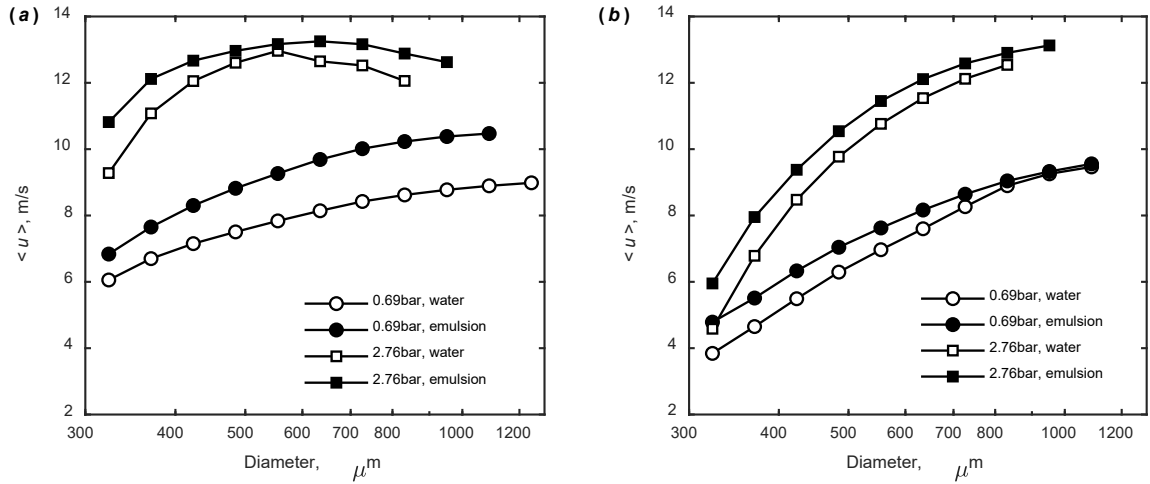

**Fig. 9.** Effects of oil-in-water emulsion and pressure on the mean droplet axial velocity at (a) the spray center ($y = 0$), and (b) the spray edge ($|y| = 20$ cm for 0.69 case and $|y| = 30$ cm for 2.76 bar case). All emulsions are 0.1% oil by volume.

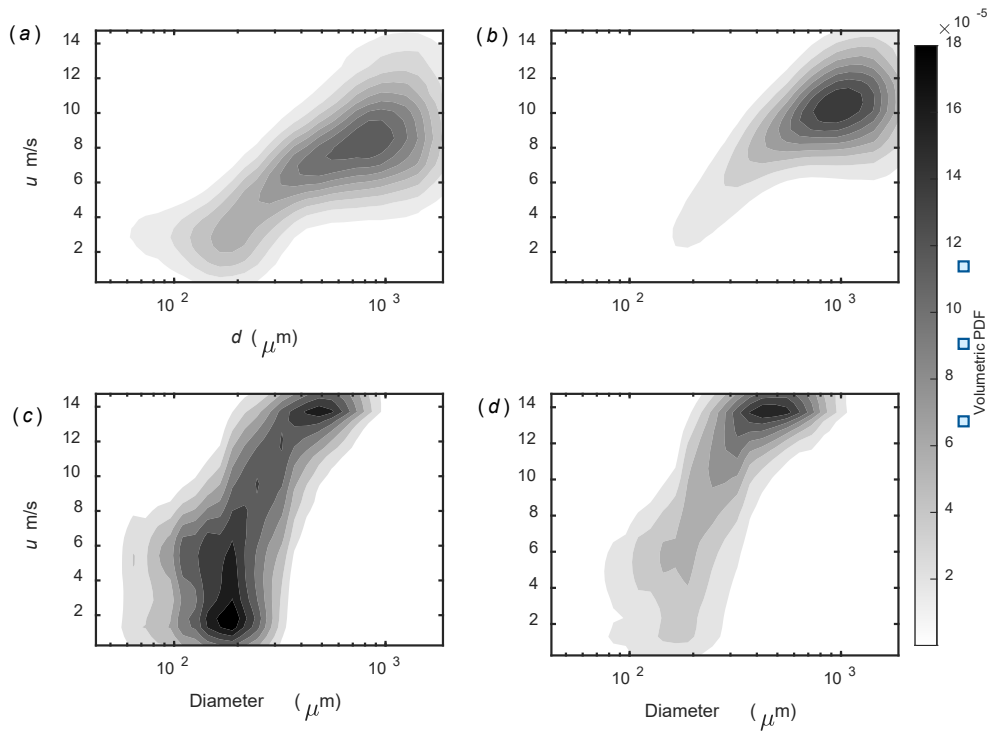

**Fig. 10.** Joint probability density functions in diameter and horizontal speed at the spray center ($y = 0$) for (a) 0.69 bar, water spray, (b) 0.69 bar, 0.1% oil-in-water emulsion spray, (c) 2.76 bar, water spray, and (d) 2.76 bar, 0.1% oil-in-water emulsion sprays.

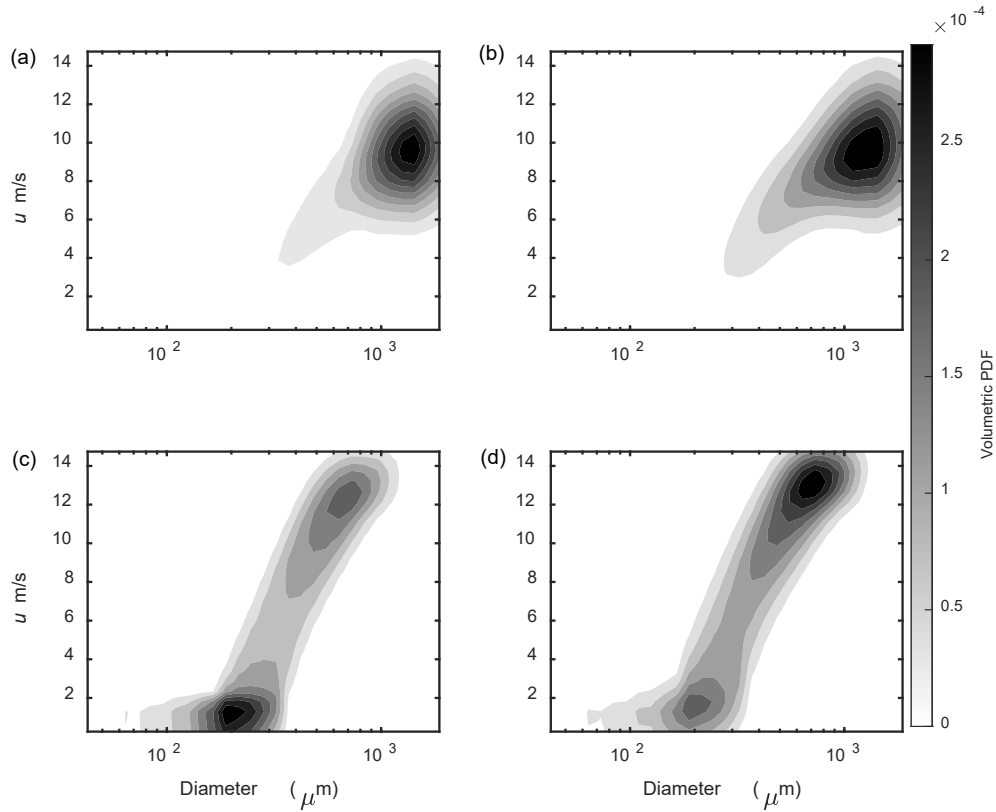

**Fig. 10.** Joint probability density functions in diameter and horizontal speed at the spray edge ($y$ = 20 cm at 0.69 bar and $y$ = 30 cm at 2.76 bar) for (a) 0.69 bar, water spray, (b) 0.69 bar, oil-in-water emulsion spray, (c) 2.76 bar, water spray, and (d) 2.76 bar, oil-in-water emulsion spray.

As expected, the pressure increase causes the mean axial velocity to increase for all cases at all size ranges, yet the magnitude of the increase differs between cases. Oil-in-water emulsion droplets have increased mean axial velocities in all conditions. This is presumably because of reduced drag influences on the sheet prior the breakup, less energy dissipation during breakup (leading to more spherical droplets), and a less efficient breakup process (suppressed smaller droplet formation, hence reduced surface energy generation rates). Therefore, in addition to reducing driftable content, we suggest that the application of emulsion sprays in agriculture practices may increase canopy penetration, which is important in flat fan spray application to improve pesticide efficacy (Zhu et al., 2002; Zhu et al., 2004). Finally, the mean axial velocities at the edge of the spray are in general smaller than those measured at the center of the spray. This is simply due to the fact that droplets have to travel longer distance to reach at the measurement location ($x$ = 50 cm) at the edge.

Before concluding, we note that in a number of related fields, there is growing interest in utilizing joint size-velocity pdfs to better understand the dynamics of particle and droplet populations in their environments. Figs. 10 and 11 of the supporting information display joint size-velocity pdfs, similar to those reported by Roisman and Tropea (2001) for sprays under the same conditions and at the same locations as discussed in Figs. 7 and 9, respectively. In general, droplet axial speeds span from 1 m s$^{-1}$ to 14 m s$^{-1}$, with a linear-log scaling for the peak axial speed evident in each pdf with diameter. In joint pdfs we do observe a significant increase of small droplets volume fraction with lower velocities, and in some instances, a local peak in the joint-pdf is evident (e.g., Fig. 10c and Fig. 11c-d). The effect of addition of oil-in-water emulsion is to remove or dampen small droplets and small velocity contribution. However, in large part, the mean speed displayed in figure 9 capture the differences observed in joint size-velocity pdfs, and the pdfs simply reveal distributions in velocity at all observed droplet sizes and locations.

## 4. Conclusions

Using shadowgraphic imaging and digital inline holography (DIH), this study provides a systematic investigation of the effects of oil-in-water emulsions on the spray characteristics and resulting droplet size distributions in a flat-fan nozzle. Specifically, shadowgraphic imaging demonstrates that the addition of oil-in-water emulsion makes the ligament formation during spray less organized and thickens the ligament in comparison to those observed in single phase water sprays. Such change suppresses the formation of smaller droplets and shifts the volumetric droplet size distribution (VDSD), derived from DIH-based size measurements downstream of the breakup region, to large droplet size range. This result provides a strong support to the use of oil-in-water emulsion to reduce spray drift. Besides, we show that the varying concentration of oil is insensitive for furthering the change to the VDSD. Additionally, the VDSD exhibits a variation between the spray center and edge presumably due to the rim effects at the spray edge for all cases. Such differences are most significant for water spray at lower pressure, and less prominent for high pressure due to reduced rim thickness and oil in water cases due to the generation of rims due to hole formation in the lamella. Moreover, for the first time, we provide size dependent mean eccentricity profile for the flat fan oil-in-water emulsion sprays, which demonstrate a linear-log scaling, with conditions of pressures, locations, and oil-in-water emulsion modulate the profile. Finally, mean axial velocity profiles also show a log linear scaling for all cases, with droplets velocity span from 0 to up to ~14 m/s, and the shape of the curves determined by the drag of the corresponding droplet. The application of oil emulsion in the spray causes an increase of mean axial velocity. The observations provide quantitative evidence the addition of oil droplets are capable of dampening the wind drift portion and at the same time, increase the penetration capability of the spray.

The current study is, to our knowledge, the first to apply DIH to precisely examine spatially variable droplet size distributions for flat fan oil-in-water emulsion sprays. The result is a baseline dataset describing droplet dynamics with a single type of oil, and in the absence of flow. Future work is required to mechanistically understand how oil-in-water emulsion properties, combined with liquid supply pressure and nozzle type, can precisely affect the smaller mode of the droplet size distribution. Further instrument and data analysis method development is also required to extend the range of wind speeds and liquid supply pressures outside the ranges experiments have been conducted here, which are relevant to a wider variety of potential field conditions.

## 5. Acknowledgements

This work was supported by Winfield United. The authors thank Mr. Chase Christen and Mr. Ian Marabella for assistance with operation of the spray wind tunnel.

## 6. Supporting Information

Depth compressed images and videos of droplet trajectories and a list of the conditions for which DIH measurements were made are available online.

**Supplemental Information**

| Dataset # | Pressure/ bar | Oil vol% | y / cm | Total droplet count | Average # droplets per hologram | Maximum # of droplet per hologram | Minimum # droplets per hologram |
|---|---|---|---|---|---|---|---|
| 1 | 0.69 | 0 | -20 | 2720491 | 45.34 | 125 | 9 |
| 2 | 0.69 | 0 | 0 | 8615382 | 143.59 | 320 | 62 |
| 3 | 0.69 | 0 | 20 | 187412 | 3.12 | 25 | 4 |
| 4 | 2.76 | 0 | -30 | 1695142 | 28.25 | 227 | 8 |
| 5 | 2.76 | 0 | 0 | 16505276 | 275.09 | 529 | 9 |
| 6 | 2.76 | 0 | 30 | 661554 | 11.03 | 57 | 7 |
| 7 | 0.69 | 0.1 | -20 | 856999 | 14.28 | 81 | 2 |
| 8 | 0.69 | 0.1 | 0 | 1191278 | 19.85 | 60 | 4 |
| 9 | 0.69 | 0.1 | 20 | 343538 | 5.73 | 35 | 2 |
| 10 | 2.76 | 0.1 | -30 | 3695868 | 61.60 | 218 | 12 |
| 11 | 2.76 | 0.1 | 0 | 4975554 | 82.93 | 181 | 38 |
| 12 | 2.76 | 0.1 | 30 | 550520 | 9.18 | 36 | 2 |
| 13 | 0.69 | 0.2 | 0 | 1027945 | 17.13 | 54 | 2 |
| 14 | 0.69 | 1 | 0 | 857377 | 14.29 | 49 | 2 |

Table S1. Summary of datasets. For each data sets, a total 60,000 holograms are processed for generating droplet size and velocity statistics

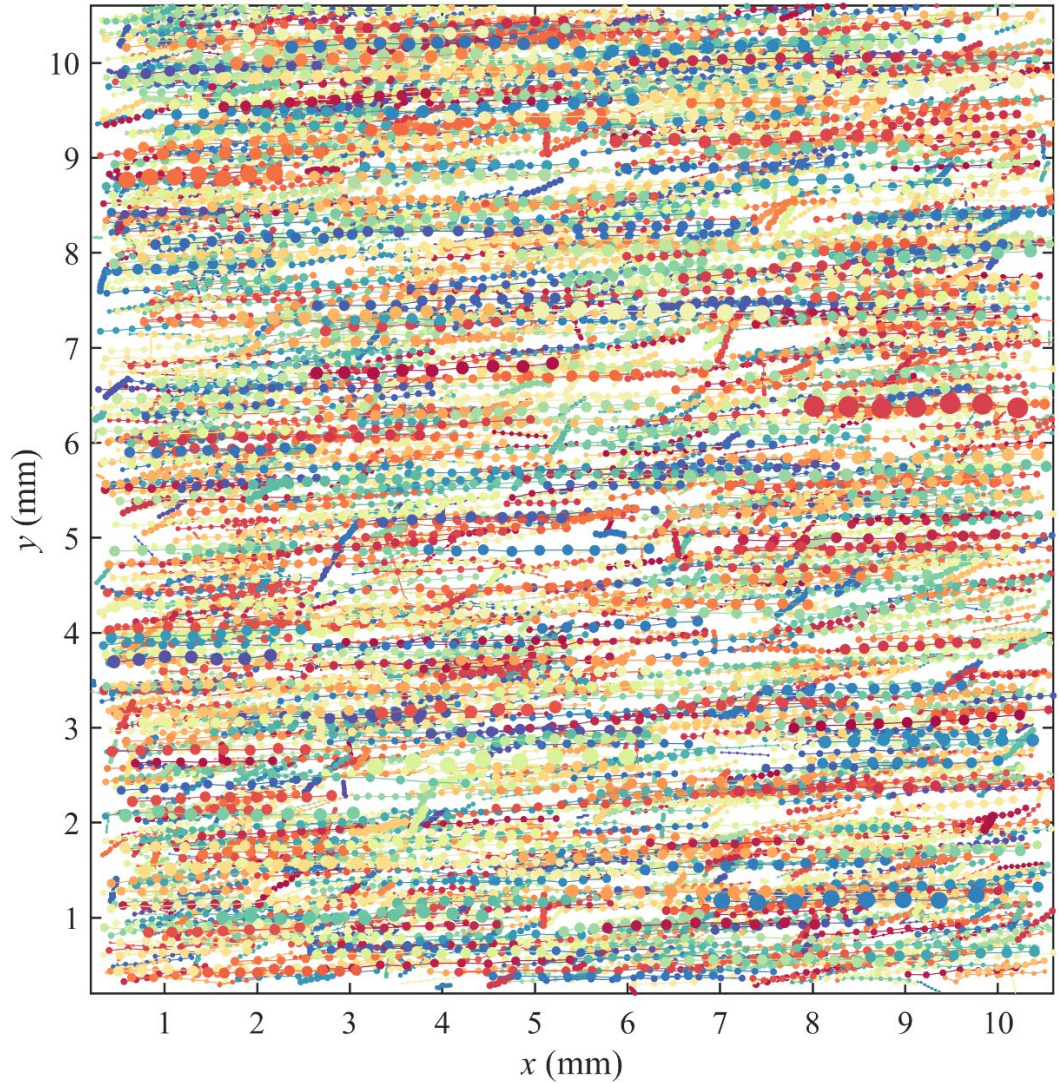

Figure S1. Sample 2D droplets trajectories reconstructed from DIH measurements, from 0.1% fish oil, 0.69 bar, at $y = 0$ dataset, with colors used to distinguish the individual tracks and marker size used to indicate relative size of the droplet.

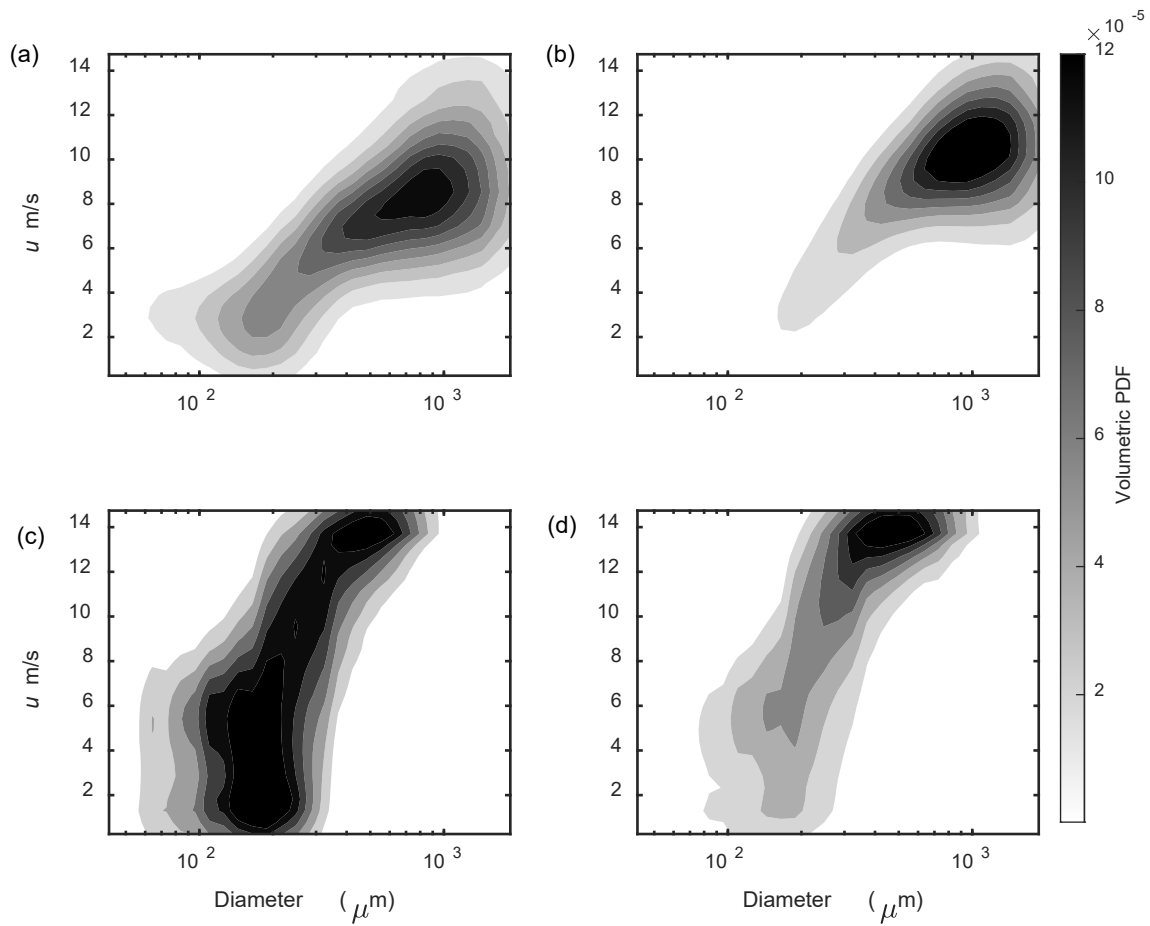

Figure S2. Joint probability density functions in diameter and horizontal speed at the spray center ($y = 0$) for (a) 0.69 bar, water spray, (b) 0.69 bar, 0.1% oil-in-water emulsion spray, (c) 2.76 bar, water spray, and (d) 2.76 bar, 0.1% oil-in-water emulsion sprays.

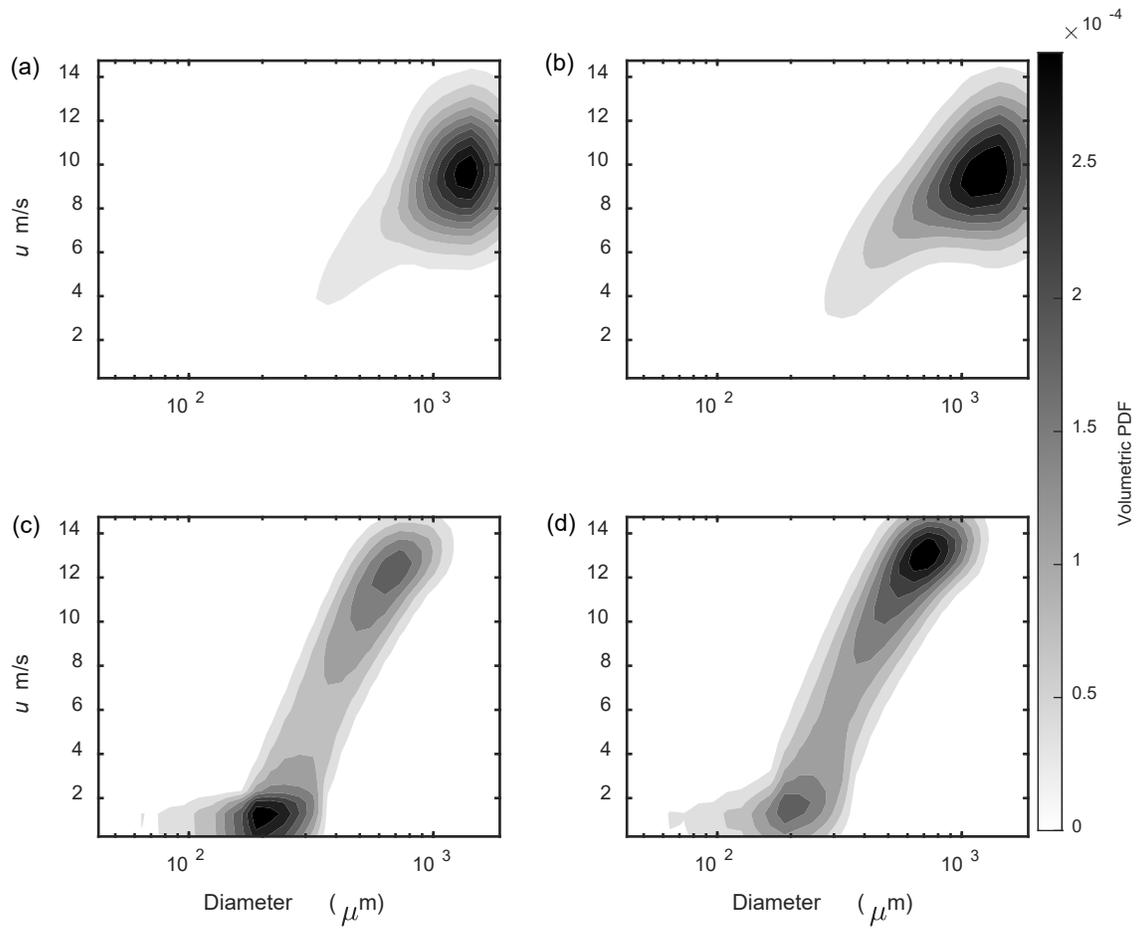

Figure S3. Joint probability density functions in diameter and horizontal speed at the spray edge ($y = 20$ cm at 0.69 bar and $y = 30$ cm at 2.76 bar) for (a) 0.69 bar, water spray, (b) 0.69 bar, oil-in-water emulsion spray, (c) 2.76 bar, water spray, and (d) 2.76 bar, oil-in-water emulsion spray.